\documentclass[%
 reprint,
 amsmath,amssymb,
 aps,
]{revtex4-2}

\usepackage{graphicx}
\usepackage{dcolumn}
\usepackage{bm}

\usepackage[version=4,arrows=pgf-filled,
textfontname=sffamily,
mathfontname=mathsf]{mhchem}

\begin{document}

\preprint{APS/123-QED}

\title{All-Dielectric Resonant Cavity Electro-Optic Transduction Between Microwave and Telecom}

\author{Mihir Khanna}
\email{mihir.khanna@pitt.edu}
\author{Yang Hu}
\altaffiliation[Present Address: ]{Skyworks Solutions Inc.~, Irvine, CA, USA}
\author{Thomas P. Purdy}
 \email{tpp9@pitt.edu}
\affiliation{
 Department of Physics and Astronomy, University of Pittsburgh, Pittsburgh, PA, USA\\
}

\date{\today}

\begin{abstract}
We present a resonant electro-optic transducer for efficient conversion between microwave and telecom wavelength photons.  Our platform employs a bulk lithium niobate crystal whose large dielectric constant creates wavelength-scale confinement of microwave photons.  By incorporating this crystal within a high-finesse Fabry--Perot optical cavity, microwave photons couple to optical photons through the electro-optic effect.  We demonstrate the ability to tune our system into triply resonant operation, where microwave photons, optical pump photons, and upconverted optical photons are simultaneously resonant with high quality factor electromagnetic modes of the system. The device achieves photon number conversion efficiency at the percent level, comparable to state-of-the-art devices at room temperature --- sufficient to resolve the thermal occupation of the microwave mode --- while avoiding the noise and loss associated with metal electrodes. These results establish our all-dielectric devices as a promising platform for high-precision sensing of optically detected microwave fields and as a viable route toward single-photon-level microwave--optical quantum transduction.
\end{abstract}

\maketitle

\section{Introduction}
Efficient conversion between microwave and optical photons is a key enabling technology for quantum information transduction in tasks where individual photons matter~\cite{Lauk2020,Lambert2020}. In efforts to interconnect superconducting quantum processors, high-fidelity microwave–optical transduction is expected to enable the distribution of information and entanglement via room-temperature fiber-optic communication channels~\cite{Kimble2008}. Beyond networking, efficient converters will facilitate the detection of microwave fields with sensitivities approaching fundamental quantum limits, precision temperature metrology via upconverting thermal blackbody radiation, and optical manipulation of microwave devices (e.g.~laser cooling microwave modes)~\cite{Tsang2010, Aspelmeyer2014}. Several transduction mechanisms have been explored~\cite{Han2021} --- including optomechanics, atomic ensembles, and spin systems --- but these require intermediary processes that add complexity and noise and limit bandwidth.  Electro-optic (EO) transduction offers a more direct and high-bandwidth route~\cite{hu2025}. Here, microwave photons modulate optical fields in a nonlinear medium without intermediaries. However, conventional EO modulators used ubiquitously in telecommunications achieve only minuscule conversion efficiencies~\cite{Ilchenko2003,Youssefi2021}, orders of magnitude below what quantum applications demand.  Several recent EO devices aimed at quantum applications have made dramatic progress by employing superconducting electrodes to tightly confine microwave photons near nanophotonic devices~\cite{Javerzac-Galy2016,Fan2018,Mckenna2020, Holzgrafe2020, Xu2021,Warner2025}. However, this architecture has inherent drawbacks: superconducting operation precludes room temperature applications, absorbed optical photons generate excess noise~\cite{Tang2024}, and heat dissipation in nanoscale devices limits optical power handling.  Moving to larger whispering gallery mode optical resonator devices has also proven successful~\cite{Cohen2001,Ilchenko2003,Rueda2016,Hease2020}, where the reduced coupling
between microwave and optical fields  is balanced by the increased optical power handling.  Such devices have reached high efficiency by operating with short, low-duty-cycle optical pump pulses to mitigate heating~\cite{Sahu2022,Qiu2023,Sahu2023}. Here, we investigate the idea that foregoing electrodes in an EO device can improve optical power handling and reduce loss sufficiently to establish a viable route to efficient microwave--optical quantum transduction.

We present a bulk, all-dielectric approach to resonant electro-optics that exploits the large microwave dielectric constant of lithium niobate (LN) to achieve wavelength-scale confinement of microwave photons without metal electrodes, enabling efficient  transduction. Our device consists of a cm-scale LN crystal functioning simultaneously as a dielectric microwave resonator and as part of a high-finesse Fabry--Perot optical cavity (See Fig.~\ref{fig:f1}.).  The geometry is engineered to achieve triply-resonant operation, where microwave, optical pump, and output photons are all resonant with electromagnetic modes of the system.  We measure a single photon EO coupling rate of $g_{0}/2\pi=1.5\pm0.3\ \mathrm{Hz}$, and with strong continuous wave (CW) optical pumping we achieve a cooperativity (a dimensionless measure of transduction) of $C=(1.7\pm 0.8)\times10^{-2}$, on par with other state-of-the-art, room-temperature, CW  devices~\cite{Rueda2016,Hease2020,zhang2025}. Further, by pumping the microwave mode of our device, we can strongly couple the two optical modes. We measure the resultant optical normal mode splitting to directly calibrate the EO coupling rate, in good agreement with simulations. Finally, we assess the possibility of pushing all-dielectric EO devices into the strong coupling and quantum regimes of operation.

\section{Design and Simulation}

\begin{figure}
    \centering
    \includegraphics[width=\columnwidth]{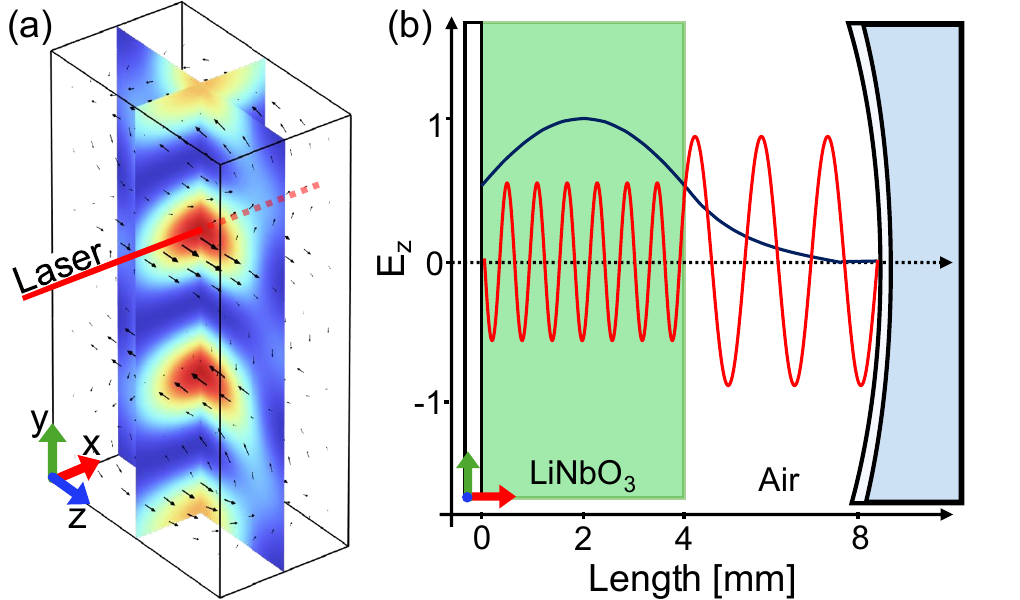}
    \caption{All-dielectric electro-optic resonator design. (a) Finite element simulation (COMSOL) of the electric field energy distribution of the  $TM_{131}$ microwave dielectric resonator mode of a $4\times12\times8\ \mathrm{mm}$ LN slab. The black arrows indicate the direction of the electric field. The red line depicts the optical beam path, which passes through the region of maximum microwave field. Here, both the optical and microwave electric fields are along the crystal $z$ axis, where the electro-optic coefficient is largest. (b) Overlap of microwave and optical modes in a Fabry--Perot cavity half-filled with LN. The $z$ component of the microwave field along the laser path is shown in dark blue. The optical cavity consists of a curved mirror and a mirror coating on the back of the LN slab. A zoomed-in segment of the optical field near the air--dielectric interface is shown in red.}
    \label{fig:f1}
\end{figure}
Resonant EO devices, so-called cavity electro-optic systems~\cite{Tsang2010,Tsang2011}, can be modeled as parametrically coupled harmonic oscillators. The Pockels effect, which is the basis of this coupling, allows a  microwave electric field to modulate the optical refractive index of a nonlinear medium (e.g.~lithium niobate) in an optical cavity, creating sidebands on optical light pumping the system.  If only one of these sidebands matches the frequency of an optical cavity mode (the output mode), its amplitude is resonantly enhanced, and the other sideband is suppressed, resulting in single-sideband modulation. The net effect is the exchange of microwave and output optical photons with the energy difference being made up by pump photons.  

The central figure of merit governing the microwave--optical transduction efficiency in EO systems is the cooperativity, 
\begin{equation}
C=\frac{4 {N}_{p}g_0^2}{\kappa_o \kappa_m}.    
\end{equation}
  Here $\kappa_{o}$ ($\kappa_{m}$) is the total decay rate of the optical (microwave) resonator, and ${N}_p$ is the average occupation of the optical pump mode.  For triply resonant operation, the peak photon number transduction efficiency is given by~\cite{Javerzac-Galy2016,Fan2018}
\begin{equation}
\eta_{peak}=\frac{4 C}{\left(1+C\right)^2} \frac{\kappa_{o,ext}}{\kappa_o} \frac{\kappa_{m,ext}}{\kappa_{m}},
\label{etapeak}
\end{equation}
where $\kappa_{o,ext}$ ($\kappa_{m,ext}$) are the coupling rates to the optical (microwave) ports. Unit photon number transduction efficiency occurs when $C=1$ (assuming critical microwave and optical coupling).

Considering resonant EO transduction as a nonlinear optical wave-mixing process, the EO coupling rate can be calculated as~\cite{Rueda2016,Ilchenko2003,boyd}
\begin{equation}
\begin{split}
   g_{0}=\chi^{(2)} \frac{n^{2}}{\sqrt{\varepsilon_{m}}} \sqrt{\frac{\hbar \omega_{m} \omega_{p} \omega_{o}}{8 \varepsilon_{0} V_{m} V_{p} V_{o}}}\  
    \int d V \psi_{m} \psi_{p} \psi_{o},
\end{split}    
\end{equation}
where $\psi_i$, $\omega_i$, and $V_i$ are the mode shape, frequency, and volume for $i\in m,o,p$, and  $\chi^{(2)}$, $n$, $\epsilon_m$ are the second-order nonlinear susceptibility, the optical index of refraction, and the microwave dielectric constant of the EO material.  Consequently, the cooperativity scales as  $C\propto\frac{Q_m}{V_m}$. We pursue a strategy to maximize $C$ by forming the microwave cavity as a dielectric resonator made from bulk LN, as an alternative to metal electrodes.  Noting the relatively high microwave dielectric constant of LN, $\epsilon_z\approx25$, $\epsilon_x=\epsilon_y\approx45$, tight microwave photon confinement in the low order modes of a wavelength scale device is possible.  For example, Fig.~\ref{fig:f1}(a) shows the TM$_{131}$ mode of our centimeter-scale slab of LN, where the indices  label the number of antinodes along the $x$, $y$, and $z$ axes, respectively. Our simulations find $V_m\approx100\ \mathrm{mm}^3$ for this mode, which is only a third of a cubic wavelength in LN, $(\lambda_m/\sqrt{\epsilon_z})^3$.  The electric field of this mode circulates in the $y$--$z$ plane, with the field pointing in the $z$ direction at the antinodes.  An $x$-propagating, $z$-polarized laser passing through a microwave antinode will have good spatial overlap with the microwave field, maximizing $g_0$, as shown in Fig.~\ref{fig:f1}.  The particular choice of geometry allows us to take advantage of the largest EO coefficient of LN, $r_{33}=31\ \mathrm{pm/V}$, with type-0 phase matching. 

We construct an optical Fabry--Perot cavity centered on this beam path by depositing a high reflectivity mirror coating on one $x$ face of the crystal, with an additional curved  mirror completing the cavity.  The light travels through the LN over a length $L_{\mathrm{LN}}$ and through air over a length $L_{\mathrm{air}}$, as it circulates in the cavity. Due to the reflection at the air--dielectric interface, the path of optical photons through the cavity becomes more complex since light can either circulate in a single half of the cavity or through the entire resonator. We determine the resonant frequencies of the cavity by using a transfer matrix approach~\cite{kharel2019}, or by numerically finding wave vectors that meet applicable boundary conditions (the field and its derivative are matched at the interface). The resulting standing wave optical modes have an electric field amplitude in the air gap that is a factor of $A$ larger than the amplitude in the LN, where $A$ is between $1$ and $n$.  A further consequence of this reflection at the interface is that the cavity has non-uniform longitudinal mode spacing (i.e. free spectral range (FSR)), which is critical to operating in the single-sideband regime.

Since the mode waist of the optical cavity ($\sim100\ \mu \mathrm{m}$) is much smaller than the microwave wavelength (in the crystal), the microwave mode varies little over the transverse optical mode profile allowing us to simplify to a one dimensional model. In this limit, $g_0$ is independent of the beam waist. Consequently, shrinking the waist to minimize the optical mode volume does not increase $g_0$.  Conversely, keeping a large waist has the practical advantage of increasing the optical power handling.   The coupling rate becomes
\begin{equation}\label{eq:g0}
\begin{split}
     g_{0}=r_{33} \frac{n^{2}}{\sqrt{\epsilon_{z}}} \sqrt{\frac{\hbar \omega_{m} \omega_{p} \omega_{o}}{8 \varepsilon_{0} V_{m} L_{p}^{eff} L_{o}^{eff}}} \ \times \\ 
     \int_{x=0}^{x=L_{\mathrm{LN}}} \psi_{m}(x) \frac{1}{2} \cos\left( n \Delta k x \right)d x,
\end{split}     
\end{equation}
with $\Delta k=\omega_{m}/c$, the difference in wave vectors between the pump and output modes (assuming triply resonant operation), which we assume is much smaller than the wave vectors of these modes. The effective cavity length of an optical mode is $L^{eff}=\frac{1}{2}L_{\mathrm{LN}}+\frac{A^2}{2n^2}L_{\mathrm{air}}$.  For the device we use in our experiment, we estimate $g_{0}/2\pi= 1.7\pm0.2\ \mathrm{Hz}$ from Eq.~\ref{eq:g0}, for the $TM_{131}$ mode, where the uncertainty originates from our knowledge of the position of the mirror relative to the slab. 

The air gap in the optical cavity has numerous beneficial functions. First, for triply resonant operation, without the gap, the EO interaction would not be phase matched, and the integral in Eq.~\ref{eq:g0} evaluates to zero, as the light experiences equal positive and negative phase shifts over a round trip as the microwave field oscillates over its full period. With the air gap, a process similar to quasi phase matching occurs; an optical photon is inside the LN for only half of each microwave period, maximizing the overlap integral. Second, the non-uniform FSR creates pairs of spectrally isolated optical modes to act as the pump and output. With a uniform FSR, EO transduction would equally populate the two optical modes on either side of the pump mode, as both EO sidebands would be resonant with different longitudinal modes of the cavity.  While this mode of operation is useful in some contexts (e.g.~EO frequency comb generation~\cite{rueda2019,Zhang2019}), for quantum signal transduction, a single output is typically required. Lastly, the air gap gives us the ability to vary the longitudinal mode spacing, by scanning $L_{\mathrm{air}}$ and/or the laser wavelength, allowing us to experimentally tune to triply resonant conditions. 

\section{Experimental Results}

\begin{figure}
    \centering
    \includegraphics[width=\columnwidth]{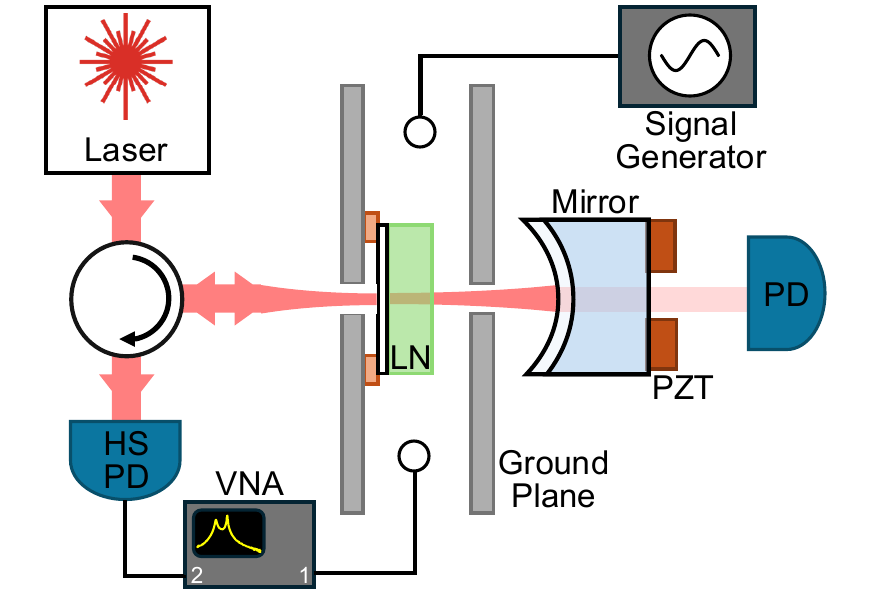}
    \caption{Simplified schematic of the experimental set-up. The LN slab (green) is affixed to an aluminum ground plane (light gray), which is mounted on a 5-axis translation stage. An opposing ground plane reduces radiative losses of the microwave modes. Microwave signals from a VNA, and from a signal generator are introduced via independent loop antennas. The optical cavity, completed by a curved mirror (light blue), is pumped by a $1550\ \mathrm{nm}$ laser, and the reflected light is collected on a high speed photodetector (HS PD). The curved mirror is attached to a ring piezoelectric actuator (brown) to scan and lock the optical cavity.  The transmitted light is collected on a photodetector (PD) as well to characterize the optical response. See Fig.~S1 for a more detailed diagram of the experimental set-up.}
    \label{fig:f2}
\end{figure}

\begin{figure*}
    \centering
    \includegraphics[width=\textwidth]{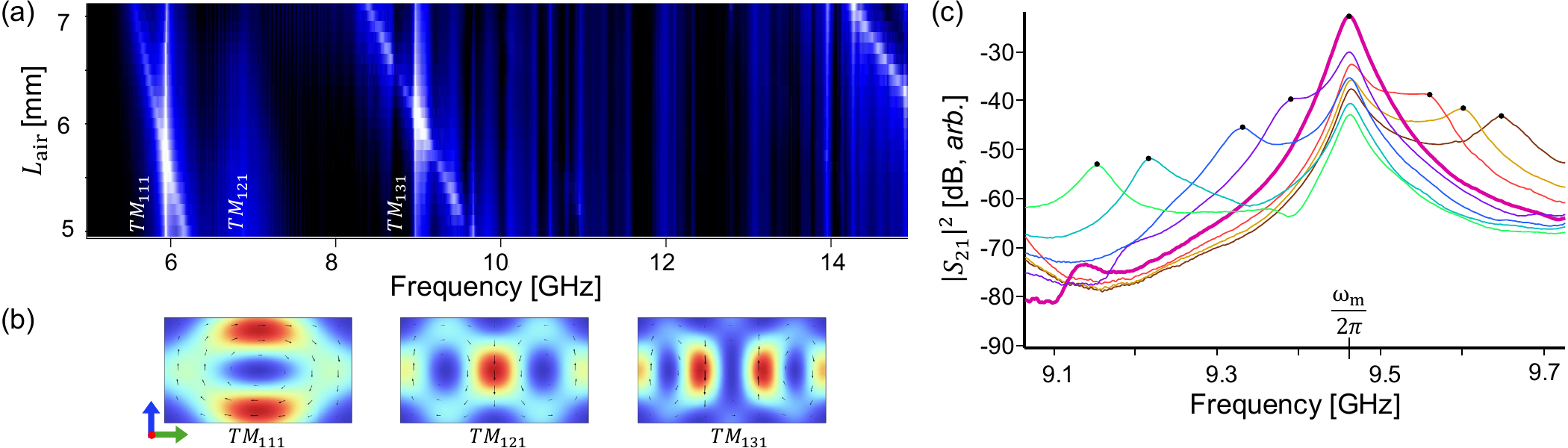}
    \caption[width=\textwidth]{Tuning into triply resonant electro-optic transduction. (a) A stacked series of $S_{21}$ traces taken as the air gap of the optical cavity is coarsely lengthened. The vertical lines correspond to the microwave modes which remain at a constant frequency. The diagonal lines correspond to the difference frequency between the pump and an output mode ($\Delta_{op}$), which vary with cavity length. The hot-spots where a diagonal line intersects a vertical line correspond to triply resonant transduction.  (b) Electric field energy distributions of three lowest order $TM$ microwave modes. The $TM_{111}$ and $TM_{131}$ modes have good spatial overlap with the optical mode, yielding strong transduction peaks. (c) $S_{21}$ traces for varying laser wavelengths while maintaining constant cavity length, to within the piezo actuator's scan range. The bold magenta trace corresponds to triply resonant conditions. The black dots represent $\Delta_{op}$ which tunes with laser wavelength. The $S_{21}$ transduction peak is maximized as $\Delta{op}$ nears $\omega_m/2\pi=9.44\ \mathrm{GHz}$ for the $TM_{131}$ mode.}
    \label{fig:f3}
\end{figure*}

The core of our experiment (See Fig.~\ref{fig:f2}) is a  $4\times12\times8\ \mathrm{mm}$ slab of LN, diced from a $4\ \mathrm{mm}$ thick, $x$-cut LN wafer coated on one side with a dielectric optical mirror.  This crystal is placed between two aluminum ground planes  separated by a  $\sim10\ \mathrm{mm}$ gap, chosen to be less than the free-space microwave wavelength to eliminate microwave radiative losses.  The aluminum has a minimal perturbation on the resonant frequency of our devices, and the losses of our system are now dominated by the dielectric loss of LN. The crystal is stood-off from one of the ground planes by $2\ \mathrm{mm}$ thick blocks of low dielectric constant, low loss polystyrene.  We couple to the low order microwave modes of the slab with a loop antenna formed on the end of a coaxial cable.  The antenna is mounted on a translation stage, which allows us to under-, over-, or critically couple to microwave modes of various spatial profiles.  The antenna is connected to port 1 of a vector network analyzer (VNA).  Microwave reflection measurements show an intrinsic quality factor of $Q_{m, int}=1.3\times10^3$  for the TM$_{131}$ mode at a frequency $\omega_m/2\pi \sim 9\ \mathrm{GHz}$.  

A curved mirror is placed outside of the ground planes (which have sub-microwave wavelength apertures for the laser) to form the optical Fabry--Perot cavity.  This mirror is mounted on a piezoelectric actuator to fine tune $L_{\mathrm{air}}$.  The slab and ground plane are mounted on a 5-axis translation stage to coarsely adjust $L_{\mathrm{air}}$ and to align the transverse location of the optical mode to the microwave antinode. For our initial set of measurements, our mirror coatings yielded an optical cavity finesse that was as high as $\sim700$, depending on the intensity of mode at the air--LN interface~\cite{kharel2019}.  A photodetector monitoring the transmitted light is used to estimate ${N}_p$.  By scanning either the cavity length or the laser wavelength, the transmission photodetector also allows us to measure the optical response function of our system.

We excite our system with a few milliwatts from a $1550\ \mathrm{nm}$ wavelength laser.  A small fraction of the reflected light is split off as a monitor of the laser--cavity detuning. This signal is used to actively maintain resonance by adjusting $L_{\mathrm{air}}$ via the piezoelectric actuator. Most of the light reflected from the cavity is collected by a high-speed photodetector, which is connected to port 2 of the VNA. When the antenna is driven with a few milliwatts of microwave power, the interference between the reflected pump and the light transduced into the output mode generates a microwave beating signal at the high-speed photodetector.  Figure~\ref{fig:f3}(a) shows how the EO response changes as $L_{\mathrm{air}}$ is coarsely tuned over a few millimeters, while the feedback control finely adjusts the length to ensure the pump remains resonant.  Two prominent peaks in the response are visible at around $6\ \mathrm{GHz}$ and $9\ \mathrm{GHz}$, independent of cavity length, corresponding to the drive frequency being resonant with the TM$_{111}$ or TM$_{131}$ microwave modes, respectively.  Two other transduction peaks that shift lower in frequency as $L_{\mathrm{air}}$ increases are also visible, traveling diagonally across the figure.  These peaks occur where the difference in frequency between the pump mode and an adjacent longitudinal optical mode is equal to the drive frequency.  Due to our cavity geometry, the longitudinal mode spacing is not uniform, and these peaks are not degenerate. When a peak corresponding to the frequency difference between optical modes becomes degenerate with a peak corresponding to  a microwave resonance, the triply resonant operation condition is met, and the EO transduction is maximized.   With the coarse cavity length fixed, we can also tune into triple resonance by adjusting the pump laser over a frequency range on the order of the cavity longitudinal mode spacing and allowing the feedback control to fine tune the cavity length, as demonstrated in Fig.~\ref{fig:f3}(c). 

\begin{figure}
    \centering
    \includegraphics[width=1\columnwidth]{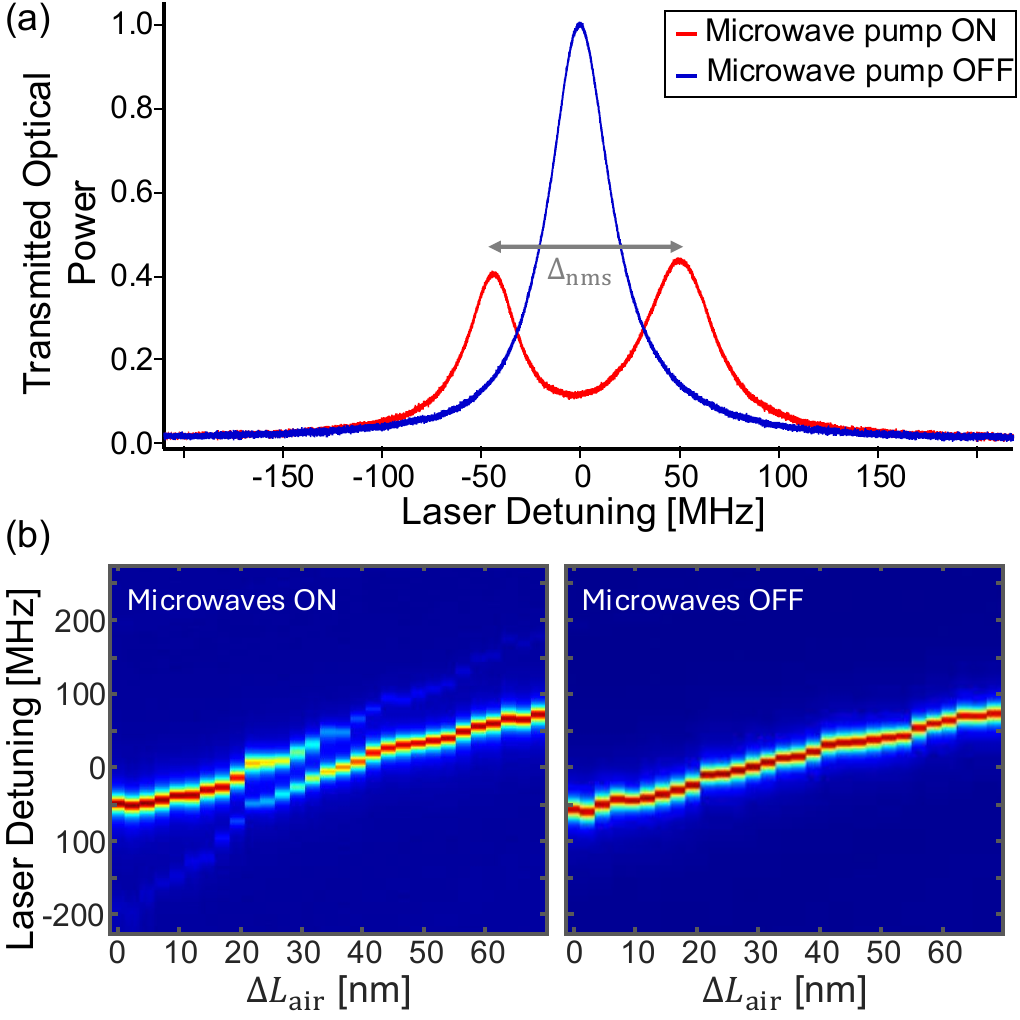}
    \caption{Strong electro-optic coupling between a pair of optical modes. (a) Optical normal mode splitting observed at triply resonant operation, when pumped with a strong microwave tone (red), compared to the Lorentzian optical response in the absence of microwave pumping (blue). The normal mode splitting $\Delta_{nms}=2\sqrt{{n}_{m}}g_0$ is used to calibrate $g_0$. (b) By finely stepping $L_{\mathrm{air}}$ while scanning the laser detuning and measuring the transmitted optical power, we tune through triple resonance, and map out the avoided crossing when a strong microwave pump is present.}
    \label{fig:f4}
\end{figure}

Next, we add an additional microwave antenna to the system to strongly pump the microwave mode.  This antenna is critically coupled to the TM$_{131}$   mode and is driven by a high power ($31\ \mathrm{dBm}$) microwave tone at $\omega_m$.  We measure the optical response function of the system by scanning the frequency of the pump laser and recording the power transmitted through the cavity.  In the absence of a strong microwave drive, we observe the Lorentzian response characteristic of an optical resonance of a Fabry--Perot cavity.  With strong microwave driving, the optical response splits into two peaks separated by about $100\ \mathrm{MHz}$, as shown in Fig.~\ref{fig:f4}(a).  We interpret this response as a normal mode splitting, where the optical pump and output modes are strongly coupled via the electro-optic interaction and the strong parametric driving at the microwave frequency (which is the difference frequency between the two optical modes)~\cite{rueda2019}. At either of these peaks, the transmitted light consists of equal amplitudes at the pump and output mode frequencies.  Our low bandwidth transmission detector records only the average transmitted power and is insensitive to the microwave beat frequency of the two optical tones.  In Fig.~\ref{fig:f4}(b) we tune our system through triple resonance by scanning $L_{\mathrm{air}}$ over a few 10's of nm.  The optical transmission curves map out the avoided crossing between the pump and output modes.   At triple resonance, the normal mode splitting is given by $\Delta_{nms}=2\sqrt{{n}_{m}}g_0$. Thus, the normal mode splitting can be used to calibrate the EO coupling. From our measured peak splitting, $\Delta_{nms}/2\pi=103\ \mathrm{MHz}$ and the measured microwave occupation, ${n}_m=(1.3\pm0.5)\times10^{15}$, we infer $g_0/2\pi=1.5\pm0.3\ \mathrm{Hz}$.  This value is in good agreement with the value predicted by our finite element simulations using Eq.~\ref{eq:g0}.

\begin{figure}
    \centering
    \includegraphics[width=\columnwidth]{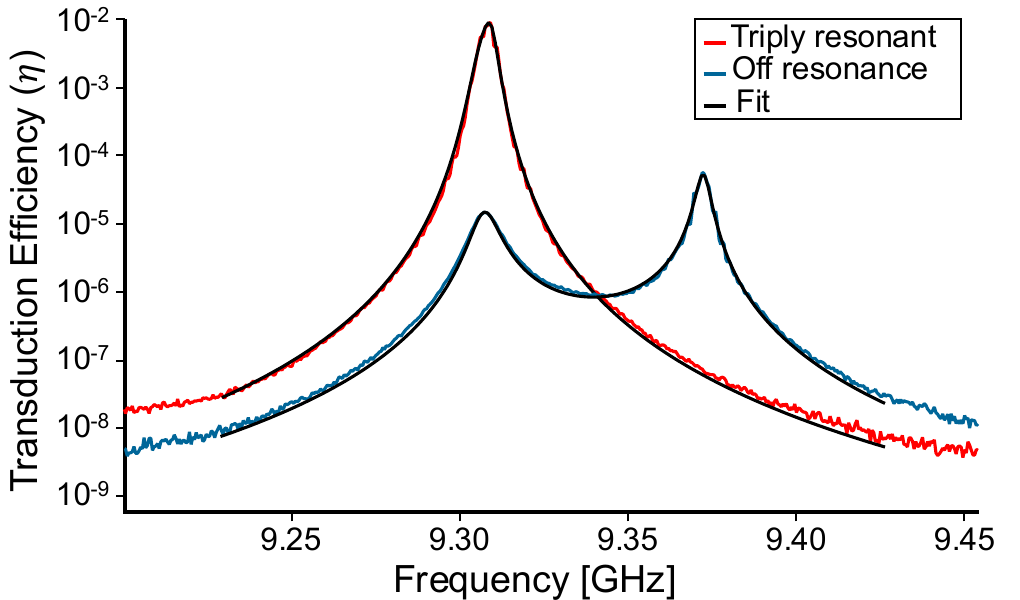}
    \caption[width=\columnwidth]{Calibrated microwave-to-optical transduction for the $TM_{131}$ mode at $9.302\ \mathrm{GHz}$. Fit to the off-resonant data (blue) yields $\kappa_m=8.54\ \mathrm{MHz}$ and $\kappa_o=4.10\ \mathrm{MHz}$. The system is pumped with $50\ \mathrm{mW}$ of optical power. For $g_0/2\pi=1.5\ \mathrm{Hz}$ this gives $C=0.017$ and $\eta_{peak}\approx1\%$ for the triply resonant data (red).}
    \label{fig:f5}
\end{figure}

Finally, we investigate the microwave-to-optical transduction efficiency of our EO system, with strong optical pumping (and the strong microwave pump removed). To this end, we constructed a new device of similar dimensions as used in previous investigations, but with higher reflectivity mirrors, aiming to decrease $\kappa_o$ and increase ${N}_p$. Additionally, we deposited an anti-reflection coating on the air--dielectric interface of the slab, reducing the reflectivity at that surface from $0.14$ to $0.006$. This reduces the tuning range of the longitudinal mode spacing with laser wavelength and crystal temperature, which decreases our sensitivity to thermal drift, allowing us to more stably maintain triply resonant operation. Further, this coating reduces both specular and diffuse scattering at the interface, into higher order transverse modes of the cavity. In the initial iterations of our device, this scattering resulted in excess optical loss when a higher order transverse modes became degenerate with the pump or output mode. We were typically able to avoid these detrimental operating conditions by carefully tuning our system parameters.  The upgraded device was measured to have $\kappa_o/2\pi=4.1\ \mathrm{MHz}$ with $\kappa_{o,ext}/2\pi=2.8\ \mathrm{MHz}$. We are able to pump this device with up to $50\ \mathrm{mW}$ incident optical power, yielding ${N}_p=6.5 \times10^{10}$.  Above this incident power, our optical cavity exhibits a strong Kerr-like nonlinear response, which we  attribute to the onset of photothermal instability.  Together, these parameters yield $C=0.017\pm0.007$. From Eq.~\ref{etapeak} we estimate our peak efficiency $\eta_{peak}\sim 1\%$. In Fig.~\ref{fig:f5} we show measurements of the microwave-to-optical transduction signal both on and off the triply resonant condition.  The photon transduction efficiency can be modeled as~\cite{Tsang2011, Fan2018}
\begin{equation}\label{eq:eta_lineshape}
\eta(\omega)=\eta_{peak}\frac{(1+C)^2}{\left|C+\left(1+\frac{2 i (\Delta_{op}-\omega)}{\kappa_o}\right)\left(1+\frac{2 i (\omega_m-\omega)}{\kappa_m}\right)\right|^2},
\end{equation}
where $\Delta_{op}$ is the difference in frequency between the pump laser and output mode frequency. We fit  both data sets to Eq.~\ref{eq:eta_lineshape}, scaled by an additional fit parameter to account for the gain of our photodetection system. These data are well matched to the expected functional form over a wide frequency range. 

\section{Outlook}

At cooperativity of $\sim10^{-2}$, we have achieved $C>1/n_{th}$, where the microwave thermal occupation is $n_{th}= k_b T/\hbar \omega_m=660$ at room temperature. In this limit, microwave thermal noise is relevant to the transduction process, making it worthwhile to consider our system as a classical, low-noise microwave sensor. Further, there are several routes for future designs to increase our electro-optic cooperativity to $C>n_{th}$, allowing operation in the quantum regime as a high-fidelity single-photon-level transducer.

For our current device, even in the absence of a coherent  microwave drive, we expect  the spectrum of the photodetected signal to be dominated by up-converted thermal microwave noise over a wide band of frequencies around $\omega_m$, assuming shot noise limited  detection of the output light.   The thermal peak to shot noise ratio is estimated as $4 C n_{th} ({\kappa_{o,ext}}/{\kappa_o})\approx15\ \mathrm{dB}$, for our current experimental parameters.  Resolving the microwave thermal occupation would open the possibility of operating our device as a low noise temperature microwave receiver. Microwave signals introduced through the antenna are detected on top of the thermal- and shot- noise floor. By overcoupling our microwave antenna,  we increase the ratio ${\kappa_{m,ext}}/{\kappa_{m,int}}$. This increases  the transduction of microwave signal photons from the input port relative to the transduction of microwave noise photons originating from internal loss in the microwave resonator.   However, overcoupling also increases the overall microwave linewidth, which reduces $C$ and $\eta$, and the magnitude of the transduced signals relative to the shot noise floor.   This implies the existence of an optimal antenna coupling, where the thermal and shot noise floors become comparable, and the SNR for photodetected signals from the microwave input port is maximized.   Characterizing this microwave receiver in terms of input referred noise, the shot noise floor will add an equivalent number of photons $n_{added}^{qu}=1/\eta$,  and the internal thermal noise will add $n_{added}^{th}=\left({\kappa_{m,int}}/{\kappa_{m,ext}}\right)n_{th}$ noise photons to the signal~\cite{zhang2025}.  From these sources, we can estimate the noise temperature of our receiver $T_n=T\left({n_{added}^{qu}+n_{added}^{th}}\right)/{{n_{th}}}$.  For the current device, by optimally overcoupling the microwave port to $\kappa_{m,ext}/2\pi\approx 50\ \mathrm{MHz}$, we expect a noise temperature of about $110\ \mathrm{K}$ (equivalent to a noise figure of $1.4\ \mathrm{dB}$). To realize this goal with our current device, we will need to upgrade our optical system to achieve shot noise limited detection, by eliminating excess laser noise, and increasing the saturation limit of our detector. Further improvements in $C$ can push our system's noise floor toward quantum limits.

It should be possible to increase the cooperativity of our devices by further shrinking the microwave mode volume.  Our simple slab geometry results in wavelength scale confinement.  However, geometries developed in the context of nanophotonics that create deeply sub-wavelength confinement in dielectric structures could be scaled up and adapted to our microwave resonators.  Here, sub-wavelength bowtie-like features concentrate electric fields to small volumes near sharp dielectric tips, while mode confinement is achieved through photonic crystal or other quasi-periodic structures~\cite{Gondarenko2006,Hu2018,Albrechtsen2022, khanna2023}.  We estimate that sub-cubic mm mode volumes should be possible with millimeter scale features fabricated from a lithium niobate slab.  Alternatively, we can exploit the extremely high dielectric constants ($\epsilon>10^2$) available at microwave frequencies in crystalline materials (e.g.~TiO$_2$ or SrTiO$_3$).  Composite microwave devices where extremely high dielectric constant materials surround relatively lower dielectric constant lithium niobate can strongly confine electric fields in the low dielectric constant material, even for sub-wavelength thickness layers~\cite{Xu04,Gondarenko2008,khanna2023}.  We note that even for mm-scale microwave devices, the waists of our optical modes are sufficiently small so as to maintain high overlap between microwave and optical fields and prevent optical clipping losses at material edges.  To this end, we have demonstrated that we are able to create high finesse optical resonators in lithium niobate crystals with transverse dimensions as low as $0.5\ \mathrm{mm}$.  Thus, we believe that a factor of $10^2$ reduction in microwave mode volume is possible, which would result in devices with $C>1$. 

Operating our devices at cryogenic temperatures is a path to the quantum regime of transduction, which requires $C>n_{th}$.  Decreasing the temperature will improve our performance in two ways.  First, even at $T\sim4\ \mathrm{K}$, $n_{th}$ is reduced below 10 quanta.  Second, the microwave loss tangent of lithium niobate has been measured to decrease by about 2 orders of magnitude at cryogenic temperatures~\cite{Zorzetti2023,Goryachev2015}, facilitating an expected boost in cooperativity.  Maintaining high power handling at cryogenic temperatures is often experimentally challenging, but bulk lithium niobate cryogenic transducers have had success at low-duty-cycle pulsed optical operation to reduce the overall heat load while keeping the transduction efficiency high~\cite{Sahu2022,Qiu2023,Sahu2023}.

\section{Conclusion}

Several designs for EO transducers have been developed over a wide range of size scales and technological platforms.  Each must make trade-offs to optimize the often competing parameters that go into the cooperativity (size vs. power handling vs. loss).  Our bulk dielectric devices are on the opposite end of the spectrum from plasmonic EO devices, which achieve sub-optical wavelength confinement of optical and microwave fields with all-metal structures at the expense of loss and power handling~\cite{Messner2023} , while nanophotonic, microphotonic, and whispering gallery mode devices span the parameter space between.  We also note, that in contrast to microwave EO devices, all-dielectric devices have been the natural choice for detection via optical upconversion of higher frequencies (in the terahertz to infrared ranges), where the material properties of nonlinear crystals and the relevant wavelengths are much closer to those of optical light (e.g.~\cite{Dam2012,Suresh25}).

Here, we have demonstrated an all-dielectric, bulk electro-optic microwave--optical transducer operating at room temperature, achieving a cooperativity of $C = (1.7 \pm 0.8)\times 10^{-2}$ and an estimated peak photon-number conversion efficiency of approximately $1\%$. Efficient transduction in this platform is enabled by simultaneously achieving high optical power handling, wavelength-scale microwave confinement, phase and polarization matching, dispersion engineering, and triply resonant mode alignment. Rather than addressing these constraints independently through complex or heterogeneous fabrication, our approach integrates them within an ultimately simple, monolithic geometry that leverages the intrinsic dielectric and electro-optic properties of lithium niobate. This strategy provides a clear and extensible architecture for electro-optic systems operating as a classical platform for low-noise microwave sensing at room temperature, and as a realistic path toward single-photon-level microwave--optical quantum transduction. 

\acknowledgments
This work was supported by the Pittsburgh Quantum Institute. Work performed in the University of Pittsburgh Dietrich School Machine Shop Core Facility (RRID:SCR\_023720), Glass Shop Core Facility (RRID:SCR\_023719), and Nanofabrication and Characterization Core Facility (RRID:SCR\_05124), and services and instruments used in this project were graciously supported, in part, by the University of Pittsburgh. MK greatly appreciates the generous support of a Pittsburgh Quantum Institute Fellowship. 

\section*{Disclosures}
The authors declare no conflicts of interest.

\section*{Data Availability}
The raw data represented in the figures and calculations are available upon reasonable request from the corresponding author.

\section*{Supplemental Material}
See the Supplement for additional experimental details.

\bibliography{bib}

\end{document}

%\preprint{}
\title{Supplemental materials for: \\ 
 All-Dielectric Resonant Cavity Electro-Optic Transduction Between Microwave and Telecom}

\author{Mihir Khanna}
\email{mihir.khanna@pitt.edu}
\author{Yang Hu}
\altaffiliation[Present Address: ]{Skyworks Solutions Inc., Irvine, CA, USA}%Lines break automatically or can be forced with \\
\author{Thomas P. Purdy}
 \email{tpp9@pitt.edu}
\affiliation{
 Department of Physics and Astronomy, University of Pittsburgh, Pittsburgh, PA, USA\\
}

\date{\today}% It is always \today, today,

\maketitle
\section{Experimental Details}
\begin{figure}[h]
  \centering
  \includegraphics[width=\columnwidth]{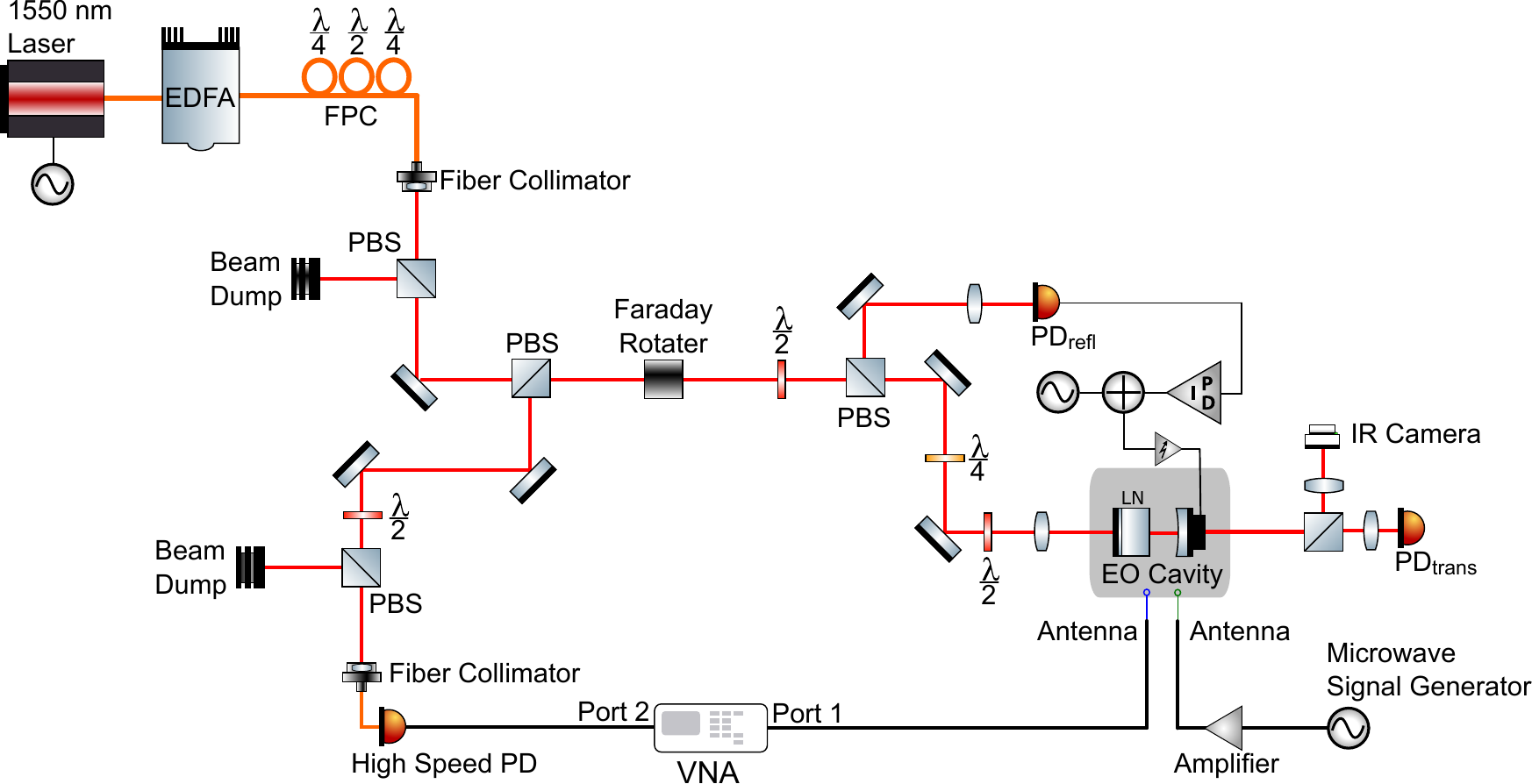}
  \caption{Detailed experimental set-up diagram. A $1550$~nm diode laser amplified by an erbium doped fiber amplifier (EDFA) pumps our electro-optic system. The laser has a modulation port which enables us to scan the laser's frequency. A fiber polarization controller (FPC) allows us to set the polarization of the light, before converting it to a free space beam. The laser is sent through an optical circulator consisting of input and output polarizing beam splitters (PBS) and a Faraday rotator.  The laser polarization is set by a quarter waveplate ($\frac{\lambda}{4}$) and half waveplate ($\frac{\lambda}{2}$) before being mode matched to the EO cavity. A small fraction of the reflected light is split off to a photodetector (PD$_{\mathrm{refl}}$) for active stabilization of the cavity length. The transmitted light from the EO cavity is collected on another photodetector (PD$_{\mathrm{trans}}$) to study the optical response of the cavity, as well as on an infrared (IR) camera to ensure the laser is coupled to the fundamental transverse mode of the Fabry--Perot cavity. Most of the light reflected from the EO cavity is directed to a high speed photodetector, which is connected to port 2 of the vector network analyzer (VNA). Port 1 of the VNA is connected to a loop antenna which is coupled to the microwave mode. A second loop antenna coupled to the microwave mode is fed by an amplified microwave signal generator, resonant with the microwave mode. The red lines depict the free-space optical beam paths, the orange lines depict fiber optics, and the black lines depict electrical and microwave connections.}
  \label{fig:setup}
\end{figure}

Our electro-optic (EO) device is a Fabry--Perot optical cavity consisting of a $4\times12\times8$~mm slab of lithium niobate (LN), diced from a 4~mm thick, $x$-cut wafer from Precision Micro-Optics Inc. with surface roughness less than than 1~nm.  The 4~mm depth of cut requires the use of a dicing blade with a larger than typical exposure, resulting in chipping within $\sim50\ \mu\mathrm{m}$ of the edges of slab and  micrometer-scale surface roughness. The surface roughness scattering of microwaves at this scale is negligible.  Prior to dicing, the LN wafer is coated on one side with a high reflectivity (HR) dielectric mirror with a measured transmission of  $0.16\%$.  The other side of the wafer is coated with an anti-reflection (AR) coating consisting of 300~nm of fused silica, which reduces the reflectivity of the surface from 0.14 to 0.006 for 1550~nm wavelength light.  Both the AR and HR coating are deposited via ion beam sputtering by Evaporated Coatings Inc.  For the cooperativity and efficiency measurements shown in Fig.~5 of the main text, the Fabry--Perot cavity was completed with a 100~mm radius of curvature mirror from LAYERTEC GmbH, with a measured transmission of $0.02\%$.  We infer the total round trip loss in the cavity to be approximately $0.1\%$, given the measured finesse of 2200.  

Our device is pumped optically with a 1550~nm laser (RIO Orion) amplified to 100~mW with an erbium doped fiber amplifier (EDFA).  As shown in Fig.~\ref{fig:setup}, this light passes through an optical circulator and is then mode matched to the cavity with around 90\% coupling efficiency.  Because of the birefringence of LN, the polarization modes of our cavity are non-degenerate, and we observe two families of longitudinal modes of orthogonal polarization, with the modes polarized along the LN $z$ axis showing the largest EO transduction as anticipated.  We intentionally couple a small fraction of the input light into the orthogonal polarization of the cavity through a combination of a quarter- and half- waveplate before the cavity.   The waveplates are configured such that a small fraction of the light reflected from the cavity is rejected by the output polarizing beam splitter of the optical circulator.  This light is directed to a photodetector, and we observe a dispersive lineshape on top of a constant background when the laser is swept across the cavity resonance due to polarization rotation induced by the birefringent cavity, a modified version of the Hansch-Couillaud locking scheme~\cite{Hansch1980} .  We use this dispersive response as an error signal to actively stabilize the cavity length to be resonant with the laser.  The photodetector signal is  fed into a Liquid Instruments Moku:Go FPGA PID controller which actuates the cavity length with a piezoelectric transducer that displaces the curved mirror.  The majority of the reflected light is directed to a high-speed photodetector (Thorlabs RXM25AF), whose output is sent to port 2 of the vector network analyzer (VNA) (Keysight P9374A).  We capture the small fraction of the light transmitted through the cavity on a 300~kHz bandwidth photodetector to monitor the optical response function and calibrate the cavity photon occupation.

The microwave modes are addressed with two few-millimeter-diameter loop antennas near the crystal.  One antenna is driven by port 1 of the VNA.   The other antenna is used only for the experiment detailed in Fig.~4 of the main text where a strong microwave pump is applied.  This antenna is driven by a microwave signal generator (SignalCore SC5511B) amplified by a 3~Watt microwave amplifier (Mini-Circuits ZVE-3W-183+).  For these measurements, the microwave signal generator is pulsed on for the duration of every tenth scan of the laser wavelength to limit microwave heating effects and to facilitate comparison of the optical response with and without strong microwave pumping on a time scale shorter than the drift in our system.  Here, because we are operating the system without active feedback stabilization of the cavity length, drifts corresponding to a few nanometers are visible as slight offsets between sequential data traces in Fig.~4.

\bibliography{bib}